\documentstyle[12pt]{article}

\font\mybb=msbm10 at 12pt
\def\bb#1{\hbox{\mybb#1}}

\newlength{\extraspace}
\newlength{\extraspaces}
\begin{document}

\addtolength{\baselineskip}{.8mm}

\thispagestyle{empty}

\begin{flushright}
{\sc OUTP}-97-74P\\
cond-mat/9712233\\
\hfill{  }\\
December 1997
\end{flushright}
\vspace{.3cm}

\begin{center}
{\Large\bf Dual Response Models \\for the Fractional Quantum Hall
Effect}\\[15mm]

{\sc L. Cooper, I.I. Kogan, A. Lopez, and R.J. Szabo} \\[2mm]
{\it Department of Physics -- Theoretical Physics\\ University of Oxford\\ 1
Keble Road, Oxford OX1 3NP, U.K.} \\[15mm]

\vskip 1.0 in

{\sc Abstract}

\begin{center}
\begin{minipage}{14cm}

It is shown that the Jain mapping between states of integer and fractional
quantum Hall systems can be described dynamically as a perturbative
renormalization of an effective Chern-Simons field theory. The effects of
mirror duality symmetries of toroidally compactified string theory on this
system are studied and it is shown that, when the gauge group is compact, the
mirror map has the same effect as the Jain map. The extrinsic ingredients of
the Jain construction appear naturally as topologically non-trivial field
configurations of the compact gauge theory giving a dynamical origin for the
Jain hierarchy of fractional quantum Hall states.

\end{minipage}
\end{center}

\end{center}

\vskip 0.5 in

\begin{flushleft}
\end{flushleft}

\vfill
\newpage
\pagestyle{plain}
\setcounter{page}{1}

Theoretical descriptions of the fractional quantum Hall effect have focused in
large part on many different hierarchy schemes \cite{hier,jain}. The most
promising one is the Jain hierarchy \cite{jain} which contains most of the
experimentally observed filling fractions. The Jain model starts with a
two-dimensional gas of electrons in an external transverse magnetic field $b$
at filling fraction
\begin{equation}
\nu^{\rm (J)}\equiv\frac{2\pi n_e}b=\frac p{2mp+1}
\label{nuJ}\end{equation}
in natural units, where $n_e$ is the electron density and $p,m$ are integers.
It then pins $2m$ units of magnetic flux to each electron. The resulting gas of
``composite" fermions ({\it i.e.} bound states of electrons and flux tubes) see
on average an effective magnetic field $B_{\rm eff}=b-4\pi mn_e$. From this it
follows that the effective filling fraction for the composite fermion moving in
the effective magnetic field is $\nu_{\rm eff}=2\pi n_e/B_{\rm eff}=p$, which
relates the fractional quantum Hall states with filling fractions (\ref{nuJ})
to integer quantum Hall states of the composite object.

Effective field theories for fractional quantum Hall systems are provided by (1
+ 1)-dimensional conformal field theories \cite{edge}, and
(2 + 1)-dimensional theories with
a coupling of the electric charge current to an additional fictitious gauge
field whose dynamics are governed by the Chern-Simons action
\cite{bulk}.
The former quantum field theories describe edge excitations
while the latter ones, which are well-known to be
equivalent to a large class of two-dimensional conformal field
theories \cite{csconf}, describe bulk dynamics of the sample.

In unrelated developments it has been realized in the past few years that
certain duality symmetries of compactified string theory \cite{dualrev} have
crucial consequences for the structure of spacetime implied by superstring
theory. A duality in string theory is a quantum symmetry that typically relates
a geometry of the target space in which the strings live to a (classically)
inequivalent one. In this letter we shall investigate the effects of certain
duality transformations on quantum Hall systems, given the relation of
Chern-Simons gauge theory to both string theory and the fractional quantum Hall
effect. The application of duality symmetries in this context has also been
discussed previously in \cite{bal} and \cite{cks}. We will apply the so-called
mirror transformation, as described in \cite{cks}, to a {\it compact}
Chern-Simons theory appropriate to the Jain hierarchy. We show that
the Jain hierarchy can be interpreted as a perturbative renormalization of
the effective field theory of an integer quantum Hall system. We then show that
the mirror transform of the Jain hierarchy can be similarly understood,
provided
that one properly accounts for extra non-perturbative instanton-induced
processes that arise due to the compactification of the gauge
group. We will see that the mirror
map takes an integer quantum Hall system with filling fraction $m$ into the
fractional one given by the Jain hierarchy (\ref{nuJ}) in which the
integers $m$ and $p$ are interchanged. The mirror map is in this way equivalent
to the Jain mapping between integer and fractional quantum Hall states.
Moreover, the non-perturbative degrees of freedom give an explicit realization
of the flux tubes which are added to electrons in the Jain construction as the
monopole-instanton field configurations of the compact Chern-Simons gauge
theory. From this point of view the flux tubes are necessary intrinsic
components of the effective field theory, so that the mirror map provides a
natural dynamical origin for the ingredients of the Jain transformation. These
dynamical properties thus give an explicit physical realization of the
geometrical phenomenon of target space duality in string theory which could be
amenable to experimental verification.

The most general quantum Hall states are described by the effective field
theory \cite{edge}
\begin{equation}
S_Q=S_{\rm CS}(K)+\sum_I\int d^3x~
\frac{Q_I}{2\pi}\,\epsilon^{\mu\nu\lambda}a_\mu\partial_\nu
A_\lambda^I
\label{qhe}\end{equation}
where
\begin{equation}
S_{\rm CS}(K)=\sum_{I,J}\frac{K_{IJ}}{4\pi}\int
d^3x~\epsilon^{\mu\nu\lambda}A^I_\mu\partial_\nu A^J_\lambda
\label{CSactionp}\end{equation}
is the Chern-Simons action for a $U(1)^p$ gauge theory with fields
$A^I$, $I=1,\dots,p$. The state described by (\ref{qhe}) contains ${\rm
rank}(K)$ quasi-particle excitations which have conserved topological current
densities $J^\mu_I=\frac1{2\pi}\epsilon^{\mu\nu\lambda}\partial_\nu
A_\lambda^I$ whose associated global conserved charge is the magnetic flux of
the gauge field $A^I$. The external electromagnetic vector potential $a$
minimally couples, with charges $Q_I$, to each of these currents. By
integrating over $A^I$ one can prove that the filling fraction for this state
is given by
\begin{equation}
\nu=\mbox{$\sum_{I,J}$}\,Q_I(K^{-1})^{IJ}Q_J
\label{nuK}\end{equation}
where we have used the fact that the filling fraction is related to the
Hall conductance $\sigma_{H}$ through $\sigma_{H}= {\nu \over {2\pi}}$.
Different hierarchy schemes of quantum Hall states correspond to different
forms of the Chern-Simons coefficient matrix $K$ and the charges $Q_I$
\cite{edge}. For the Jain hierarchy (\ref{nuJ}), one takes $Q_I=1~~\forall
I=1,\dots,p$ ({\it i.e.} all of the Chern-Simons fields carry the charge of an
electron), and the $p\times p$ symmetric matrix
\begin{equation}
K^{\rm (J)}=\pmatrix{1&0&\dots&0\cr0&1&\dots&0\cr &\dots&\dots&
\cr0&0&\dots&1\cr}+2m\pmatrix{1&1&\dots&1\cr1&1&\dots&1\cr &\dots&\dots&
\cr1&1&\dots&1\cr}
\label{jaink}\end{equation}

We will now describe how the matrix (\ref{jaink}) can be derived dynamically as
a perturbative renormalization of the Chern-Simons action. The effective theory
(\ref{qhe}) for $\nu_{\rm eff}=p$ is described by a bare coefficient matrix
$(K_0)_{IJ}=\delta_{IJ}$. Suppose now that we couple a Fermi field $\psi$ to
the fields $A^I$ with charges $Q_I$, and to an external
electromagnetic field  $a$ such that the fermions fill $2m$ Landau
levels of this external field. This system is described by the
action
\begin{eqnarray}
{\cal S}^{(R)}&=&S_{\rm CS}(K_0)+\int
d^3x~\left[\psi^\dagger\,i\left(\partial_0-ia_0-i\,\mbox{$\sum_I$}\,Q_IA_0^I-\mu
\right)\psi\right.\nonumber\\& &\left.~~~~~~~~~~~~~~~~~~~~~~~~~-
\frac1{2m_e}\,\psi^\dagger\left(\nabla-ia-i\,\mbox{$\sum_I$}\,
Q_IA^I\right)^2\psi\right]
\label{responseaction}\end{eqnarray}
where $\mu$ is the chemical potential. We shall call the fictitious fermions
created by $\psi$ ``response particles"
and the theory defined by (\ref{responseaction}) a ``response model". The
response model describes, from a dynamical point of view, the linear response
of these fermions under the mappings between quantum Hall systems in the Jain
formalism. It is an exact perturbative description that is complimentary to the
Jain model (which is non-perturbative in origin).

It is well-known that non-relativistic fermions in the presence of an
external magnetic field filling an integer number of Landau levels of
this field   renormalize the Chern-Simons
coefficients, and that this renormalization is exact at one-loop order of
perturbation theory \cite{1loop}. The bare gauge field propagator in momentum
space and in the Landau gauge is
\begin{equation}
G^{IJ}_{\mu\nu}(p)\equiv\left\langle A^I_\mu(p)A^J_\nu(-p)\right\rangle_0
=-2\pi(K_0^{-1})^{IJ}\,\epsilon_{\mu\nu\lambda}p^\lambda
\label{bareprop}\end{equation}
The only contribution to the polarization tensor $\Pi^{\mu\nu}_{IJ}(p)$
\cite{1loop} is from the fermion matter loop which can be represented
symbolically by the Feynman diagram
\unitlength=1.00mm
\linethickness{0.4pt}
\begin{equation}
\begin{picture}(100.00,11.50)
\put(0.00,5.00){\makebox(0,0)[l]{$\Pi^{\mu\nu}_{IJ}(p)~=~$}}
\thinlines
\put(22.00,5.00){\line(1,0){10.00}}
\put(42.00,5.00){\line(1,0){10.00}}
\put(26.00,8.00){\makebox(0,0)[l]{$Q_I$}}
\put(43.00,8.00){\makebox(0,0)[l]{$Q_J$}}
\put(22.00,3.00){\makebox(0,0)[l]{$\mu$}}
\put(50.00,3.00){\makebox(0,0)[l]{$\nu$}}
\put(32.00,5.00){\circle*{1.50}}
\put(42.00,5.00){\circle*{1.50}}
\thicklines
\put(37.00,5.00){\circle{10.00}}
\put(54.00,5.00){\makebox(0,0)[l]{$~=~-\frac
1{\Phi^{(R)}}\,Q_IQ_J\,\epsilon^{\mu\nu\lambda}p_\lambda$}}
\end{picture}
\label{vacpol1}\end{equation}
where $\Phi^{(R)}=\frac\pi m$ is the magnetic flux carried by the response
fermions which occupy $\nu^{(R)}=2\pi/\Phi^{(R)}=2m$ Landau levels of
the external field $a$. Here
and in the following we shall ignore the longitudinal component of
(\ref{vacpol1}) which is irrelevant to the present analysis. The polarization
tensor renormalizes the propagator (\ref{bareprop}) as
$(G^{-1})^{\mu\nu}_{IJ}(p)\to(G^{-1})^{\mu\nu}_{IJ}(p)+\Pi^{\mu\nu}_{IJ}(p)$
which leads to a renormalization of the coefficient matrix as
\begin{equation}
K^{\rm ren}_{IJ}=(K_0)_{IJ}+\nu^{(R)}\,Q_IQ_J
\label{renK}\end{equation}
When $Q_I=1$ the renormalized matrix (\ref{renK}) coincides with (\ref{jaink}).
Thus the Jain map can be interpreted as a renormalization of a Chern-Simons
gauge theory by fermion fields. Note that in this picture the fermions see only
the magnetic flux $\Phi^{(R)}$ arising from the Jain map. Thus the filling
fraction of the linear response model is $\nu^{(R)}=2m$, rather than the
physical one (\ref{nuJ}).

We shall now consider the effects of making the $U(1)^p$ gauge group of the
theory (\ref{CSactionp}) {\it compact}. This means that the pure gauge parts
$\theta^I$ of the gauge fields $A^I$ live on a $p$-dimensional torus $(S^1)^p$,
rather than on ${\bb R}^p$. The Chern-Simons gauge theory (\ref{CSactionp}) is
then equivalent to the two-dimensional conformal field theory of the $XY$ model
\cite{kogan1}
\begin{equation}
S_{XY}=\frac1{4\pi}\sum_{I,J}\int d^2z~K_{IJ}\partial_z\theta^I\partial_{\bar
z}\theta^J
\label{XYaction}\end{equation}
In string theory, the symmetric and antisymmetric parts of a generic matrix
$K_{IJ}$ coincide (up to a factor of $1\over {\alpha '}$ where $\sqrt{\alpha'}$
is the string length scale) with,
respectively, the graviton and antisymmetric tensor condensates.

The compactification of the gauge group has two profound consequences. The
first one is that the spectrum of the gauge theory now contains magnetic
monopole-instantons which have dramatic effects on the non-perturbative
dynamics of the system. In the Hamiltonian formalism they appear as
topologically non-trivial configurations of the compact gauge group when the
vacuum is projected onto the gauge-invariant subspace of the Hilbert space of
the quantum field theory. Associated with each of the $U(1)$ factors is the
monopole-instanton operator \cite{cks,lee,ckl}
\begin{equation}
V_I(x_0)=\exp\left\{-i\int d^2x~\left(\frac1{2\pi}\sum_J(K_{\rm
sym})_{IJ}A_i^J\,\partial_x^i\log|x-x_0|-{\rm
arg}(x-x_0)\,J_I^0(x)\right)\right\}
\label{miop}\end{equation}
which is a generator of compact (periodic) gauge transformations. Note that it
depends only on the symmetric part $K_{\rm sym}$ of the coefficient matrix. It
can be shown that the operator $V_I(x_0)^{n^I}$ creates a point-like magnetic
vortex of flux $2\pi n^I$ at $x_0\in{\bb R}^2$. By Gauss' law, it then also
carries electric charge $\Delta Q_I=2(K_{\rm sym})_{IJ}n^J$, so that the
monopole is an instanton that interpolates between topologically inequivalent
vacua of the Chern-Simons gauge theory labelled by the monopole numbers $n^I$
(the topological charge of $A^I$). Furthermore, it can be shown that
single-valuedness of the action of the operator (\ref{miop}) (as a function on
the torus) on physical (gauge-invariant) states gives the quantization
condition \cite{cks,ckl}
\begin{equation}
Q_I=q_I-\mbox{$\sum_J$}\,K_{IJ}\,n^J
\label{quantcondn}\end{equation}
on the spectrum of allowed charges $Q_I=\int d^2x~J_I^0(x)$ coupled to the
gauge theory. Here $q_I$ is an integer representing the winding number of the
particle around the $I^{\rm th}$ monopole-instanton.

The second consequence of the compactification is the existence of geometric
duality transformations which leave the conformal quantum field theory
(\ref{XYaction}) invariant. Of interest to us in this letter are the $p$ mirror
maps $\mu_J$ which act on a generic coefficient matrix as \cite{dualrev,cks}
\begin{equation}
K\to\widetilde{K}_J=\mu_J(K)\equiv\left[(I_p-E_J)K+E_J\right]
\left[E_JK+(I_p-E_J)\right]^{-1}
\label{mirrormap}\end{equation}
where $I_p$ is the $p\times p$ identity matrix and
$(E_J)_{KL}=\delta_{KJ}\delta_{LJ}$. The mirror transformation
(\ref{mirrormap}) is one of the $p$ factorized $T$-duality maps of bosonic
string theory compactified on a $p$-dimensional torus which inverts the radius
of the $J^{\rm th}$ cycle of $(S^1)^p$ while keeping the radii of all other
cycles fixed. It acts on the spectrum of allowed charges (\ref{quantcondn}) by
interchanging the $J^{\rm th}$ particle winding number $q_J$ and monopole
number $n^J$, {\it i.e.} it sends $Q_J\to-Q_J$ leaving all other $Q_I$'s
unchanged \cite{dualrev,cks}. In terms of the $XY$ model the mirror map
exchanges spin-wave and magnetic vortex degrees of freedom along the $J^{\rm
th}$ direction of the spin lattice.

Now we apply the mirror map (\ref{mirrormap}) to the first $N$ quasi-particles
in the coefficient matrix (\ref{jaink}) of the Jain hierarchy. Using induction
on $N$, we find after some algebra
\begin{eqnarray}
\widetilde{K}^{\rm (J)}_{1,\dots,N}&\equiv&[\mu_1\cdots\mu_N(K^{\rm (J)})]_{\rm
sym}\nonumber\\&=&\pmatrix{1& 0 & \dots & 0 & \dots & 0 \cr0 & \ddots & &
\vdots &\ddots & \vdots \cr\vdots & & 1& 0 & \dots & 0 \cr0 & \dots & 0 &1 &
\dots & 0 \cr\vdots & \ddots & \vdots & \vdots & \ddots & \vdots \cr0 & \dots &
0 & 0& \dots & 1}+\widetilde{\nu}^{(R)}_N\pmatrix{-1& 0 & \dots & 0 & \dots & 0
\cr0 & \ddots & & \vdots &\ddots & \vdots \cr\vdots & & -1& 0 & \dots & 0 \cr0
& \dots & 0 &1 & \dots & 1 \cr\vdots & \ddots & \vdots & \vdots & \ddots &
\vdots \cr0 & \dots & 0 & 1& \dots & 1}
\label{mirrorjaink1}\end{eqnarray}
where
\begin{equation}
\widetilde{\nu}^{(R)}_N=\frac{2m}{2mN+1}
\label{mirrornuresp}\end{equation}
and the second matrix in (\ref{mirrorjaink1}) contains an $N\times N$
block sub-matrix which is proportional to the identity,  and a
$(p-N)\times(p-N)$ block sub-matrix of 1's. Note that the matrices
determined by the transformation (\ref{mirrormap}) in general contain
an antisymmetric part. This piece is identified, from the
two-dimensional point of view, as an instanton tensor which does not
contribute to the local dynamics of the model. It only alters the
global, topological properties of the theory, and therefore it does
not affect local observables of the bulk theory such as the Hall
conductivity $\sigma_H\propto\nu$. This can be seen by calculating the
gauge propagator (\ref{bareprop}) of the general gauge theory
(\ref{CSactionp}), which is easily shown to depend only on the symmetric part
of $K$. Thus we consider only the symmetric part of the mirror matrix in
(\ref{mirrorjaink1}). The result (\ref{mirrorjaink1}) easily generalizes to a
succession of $N$ mirror maps $\mu_{J_1}\cdots\mu_{J_N}$ along $N$ independent
directions $J_k$. It always replaces, for each $k=1,\dots,N$, the $J_k$-th row
and column of the block matrix of 1's in (\ref{jaink}) by a $-1$ in the
diagonal entry and 0's everywhere else.

The mirror coefficient matrix (\ref{mirrorjaink1}) can be given a response
model interpretation, which we shall call the `dual Jain model'. Because
(\ref{quantcondn}) is derived in the Hamiltonian formalism, it is the {\it
bare} coefficient matrix that appears there. Since $\mu_J(K_0)=K_0$, we
therefore consider the spectrum of charges (\ref{quantcondn}) with coefficient
matrix $(K_0)_{IJ}=\delta_{IJ}$ appropriate to the response action
(\ref{responseaction}). In the initial Jain model, we have charges $Q_I=q_I=1$
and monopole flux units $n^I=0$. In the mirror theory described by
(\ref{mirrorjaink1}), we exchange the first $N$ particle charges with
topological charges, so that $Q_i=-n^i=-1$ and $q_i=0$ for $i=1,\dots,N$ while
$Q_I=q_I=1$ and $n^I=0$ otherwise. The response fermions see another unit of
magnetic flux coming from each of the $N$ monopole-instantons of the theory
which have $n^i=1$. In the mirror model, the total flux carried by the response
particles is thus $\widetilde{\Phi}^{(R)}=2\pi(\frac1{2m}+N)$ which leads to
the response filling fraction
$\widetilde{\nu}^{(R)}=2\pi/\widetilde{\Phi}^{(R)}$ given by
(\ref{mirrornuresp}). The mirror map in the response picture thus attaches flux
tubes to the fictitious fermions and maps the integer quantum Hall system with
$\nu^{(R)}=2m$ to the fractional one with filling fraction
(\ref{mirrornuresp}). But, modulo an overall factor of 2, the filling fractions
(\ref{mirrornuresp}) are just those of the Jain hierarchy (\ref{nuJ}) with
$p\to m$ and the flux units $m$ in the original model interpreted now
as monopole numbers. The mirror transformation thus relates integer
and fractional quantum Hall states to each other, from the response
model point of view, in exactly the same way that the Jain model of
composite fermions does, {\it i.e.} the mirror map $\mu_J$ plays the
role of the Jain map for response particles.

Next we demonstrate that the matrix (\ref{mirrorjaink1}) can be
obtained  exactly as a
one-loop renormalization in the theory (\ref{responseaction}). For each
$I,J=N+1,\dots,p$ we obtain the block structure of 1's in (\ref{mirrorjaink1})
exactly as in (\ref{renK}) from the vacuum polarization (\ref{vacpol1}) (with
$\Phi^{(R)}\to\widetilde{\Phi}^{(R)}$). When $I\in\{N+1,\dots,p\}$ and
$J=j\in\{1,\dots,N\}$, in addition to (\ref{vacpol1}) there is a
renormalization coming from the interaction of the fermion with the
monopole-instanton background. This induces an instanton transition that
changes the particle's charge from $Q_j=-1$ to $Q_j+\Delta Q_j=-Q_j$. It gives
the vacuum polarization contribution
\unitlength=1.00mm
\linethickness{0.4pt}
\begin{equation}
\begin{picture}(100.00,11.50)
\put(0.00,5.00){\makebox(0,0)[l]{$\widetilde{\Pi}^{\mu\nu}_{Ij}(p)~=~$}}
\thinlines
\put(22.00,5.00){\line(1,0){10.00}}
\put(42.00,5.00){\line(1,0){10.00}}
\put(26.00,8.00){\makebox(0,0)[l]{$Q_I$}}
\put(43.00,8.00){\makebox(0,0)[l]{$-Q_j$}}
\put(22.00,3.00){\makebox(0,0)[l]{$\mu$}}
\put(50.00,3.00){\makebox(0,0)[l]{$\nu$}}
\put(32.00,5.00){\circle*{1.50}}
\put(42.00,5.00){\circle*{1.50}}
\thicklines
\put(37.00,5.00){\circle{10.00}}
\put(35.50,10.00){\makebox(0,0)[l]{$\times$}}
\put(54.00,5.00){\makebox(0,0)[l]{$~=~+\frac
1{\widetilde{\Phi}^{(R)}}\,Q_IQ_j\,\epsilon^{\mu\nu\lambda}p_\lambda$}}
\end{picture}
\label{vacpol1mirror}\end{equation}
where the cross on the fermion loop indicates the interaction of the fermion
with the $j^{\rm th}$ monopole-instanton as it propagates through space-time.
The total vacuum polarization is the sum of (\ref{vacpol1}) and
(\ref{vacpol1mirror}), which are of {\it opposite} sign. This gives the two
$(p-N)\times(p-N)$ block matrices of 0's in (\ref{mirrorjaink1}).

Now suppose that $I,J=i,j\in\{1,\dots,N\}$ with $i\neq j$. There are then two
renormalizations of the sort (\ref{vacpol1mirror}) corresponding to the
interaction of either charge $Q_i$ or $Q_j$ with the monopole-instanton
background. In addition there is also the contribution
\unitlength=1.00mm
\linethickness{0.4pt}
\begin{equation}
\begin{picture}(100.00,11.50)
\put(0.00,5.00){\makebox(0,0)[l]{$\widetilde{\Pi}^{\mu\nu}_{ij}(p)~=~$}}
\thinlines
\put(22.00,5.00){\line(1,0){10.00}}
\put(42.00,5.00){\line(1,0){10.00}}
\put(24.00,8.00){\makebox(0,0)[l]{$-Q_i$}}
\put(43.00,8.00){\makebox(0,0)[l]{$-Q_j$}}
\put(22.00,3.00){\makebox(0,0)[l]{$\mu$}}
\put(50.00,3.00){\makebox(0,0)[l]{$\nu$}}
\put(32.00,5.00){\circle*{1.50}}
\put(42.00,5.00){\circle*{1.50}}
\thicklines
\put(37.00,5.00){\circle{10.00}}
\put(35.50,10.00){\makebox(0,0)[l]{$\times$}}
\put(35.50,0.00){\makebox(0,0)[l]{$\times$}}
\put(54.00,5.00){\makebox(0,0)[l]{$~=~-\frac
1{\widetilde{\Phi}^{(R)}}\,Q_iQ_j\,\epsilon^{\mu\nu\lambda}p_\lambda$}}
\end{picture}
\label{vacpol2mirror}\end{equation}
where the two crosses signify that {\it both} charges undergo the
instanton-induced charge non-conservation process simultaneously. The sum of
these four processes again yields no overall renormalization. However, when
$i=j$ we have only three such processes, as the simultaneous interaction
(\ref{vacpol2mirror}) does not occur for a single charge. The net contribution
coincides with that of (\ref{vacpol1mirror}), yielding the diagonal sub-matrix
of $-1$'s in (\ref{mirrorjaink1}).

Thus the proper incorporation of the sum over inequivalent vacua of the
compact gauge theory in (\ref{responseaction}) describes exactly the mirror
matrix (\ref{mirrorjaink1}) as a renormalization of the bare coefficient matrix
$K_0$. However, inverting (\ref{mirrorjaink1}), we find after some algebra that
the filling fractions (\ref{nuK}) for the mirror system are
\begin{equation}
\widetilde{\nu}_N=(1+2mN)\left(\frac N{1+2m(N-1)}+\frac{p-N}{1+2mp}\right)
\label{mirrornu}\end{equation}
which in general fall out of the range of the currently known stable fractions
for the fractional quantum Hall effect. For these  filling
fractions other effects that have not been considered so far
play an important role and should be included in the description of
the system.  For instance, in the context of our
analysis, the incorporation of spin degrees of freedom  would
involve, in the more general case, the generalization of the approach
described above to   an effective Chern-Simons field theory with compact
$SU(2)$ gauge group.

\bigskip

\noindent
{\bf Acknowledgements:} {\sc l.c.} gratefully acknowledges financial support
from the University of Canterbury, New Zealand. The work of {\sc i.i.k.} and
{\sc r.j.s.} was supported in part by PPARC (U.K.), and that of {\sc a.l.} was
supported in paper by EPSRC (U.K.).


\begin{thebibliography}{99}

\baselineskip=12pt

\bibitem{hier} R.E. Prange and S.M. Girvin, eds., {\it The Quantum Hall Effect}
(Springer-Verlag, Berlin, 1987), and references therein.

\bibitem{jain} J.K. Jain, Adv. Phys. {\bf 41}, 105 (1992).

\bibitem{edge} X.G. Wen, Adv. Phys. {\bf 44}, 405 (1995), and references
therein.

\bibitem{bulk} A. Lopez and E. Fradkin, Phys. Rev. B {\bf 44}, 178 (1991); S.C.
Zhang, Int. J. Mod. Phys. B {\bf 6}, 25 (1992), and references therein.

\bibitem{csconf} E. Witten, Comm. Math. Phys. {\bf 121}, 351 (1989); G. Moore
and N. Seiberg, Phys. Lett. B {\bf 220}, 422 (1989).

\bibitem{dualrev} A. Giveon, M. Porrati, and E. Rabinovici, Phys. Rep. {\bf
244}, 77 (1994).

\bibitem{bal} A. Shapere and F. Wilczek, Nucl. Phys. {\bf B320}, 669 (1989);
A.P. Balachandran, L. Chandar, and B. Sathiapalan, {\it ibid.} {\bf B443}, 465
(1995); Int. J. Mod. Phys. A {\bf 11}, 3587 (1996).

\bibitem{cks} L. Cooper, I.I. Kogan, and R.J. Szabo, hep-th/9710179.

\bibitem{1loop} Y.-H. Chen, B.I. Halperin, F. Wilczek, and E. Witten, Int. J.
Mod. Phys. B {\bf 3}, 1001 (1989); J.D. Lykken, J. Sonnenschein, and N. Weiss,
{\it ibid.} A {\bf 6}, 5155 (1991).

\bibitem{kogan1} I.I. Kogan, Mod. Phys. Lett. A {\bf 6}, 501 (1991).

\bibitem{lee} K. Lee, Nucl. Phys. {\bf B373}, 735 (1992); I.I. Kogan and A.
Kovner, Phys. Rev. D {\bf 53}, 4510 (1996).

\bibitem{ckl} L. Cooper, I.I. Kogan, and K.-M. Lee, Phys. Lett. B {\bf 394}, 67
(1997).

\end{thebibliography}
\end{document}